\documentclass[superscriptaddress,preprint,longbibliography,aps,nofootinbib,showpacs,12pt]{revtex4}

\usepackage{graphicx}  
\usepackage{bm}        
\usepackage{amssymb}   
\usepackage{epsfig}
\usepackage{amsmath,amsfonts,times}
\usepackage{bm}

\usepackage{epstopdf}
\usepackage[caption=false]{subfig}
\usepackage{natbib}
\usepackage{soul}
\usepackage[utf8x]{inputenc}

\begin{document}

\title{Ruppeiner geometry of isotropic Blume-Emery-Griffiths model} %

\author{R{\i}za Erdem }
\email{e-mail: rerdem@akdeniz.edu.tr}
\affiliation{Department of Physics, Akdeniz University, 07058 Antalya, Turkey }

\author{Nigar Alata} 

\affiliation{Institute of Science, Akdeniz University, 07058 Antalya, Turkey}

\affiliation{Food Safety and Agricultural Research Centre, Akdeniz University, 07058 Antalya, Turkey}

\begin{abstract}
With the aid of Ruppeiner thermodynamic metric defined on a two-dimensional phase space of dipolar ($m$) and quadrupolar ($q$) order parameters, we derive an expression for the Ricci scalar ($R$) in the isotropic Blume-Emery-Griffiths model. Temperature dependence of $R$ is investigated for various values of bilinear to biquadratic ratio ($r$). Its behavior near the continuous/discontinuous phase transition temperatures and a tricritical point is presented. It is found that in addition to the divergence singularity and finite jumps connected with the phase transitions there are field-dependent broad extrema in the Ricci scalar.
\end{abstract}

\pacs{05.70.a, 05.70.Fh, 75.10.Hk, 02.40.Ky}

\date{\today}
	
\maketitle

\section{Introduction}

The spin-1 isotropic Blume-Emery-Griffiths (BEG) model (also known as the spin-1 Ising model with bilinear (dipolar) and biquadratic (quadrupolar) pair interactions) is one of the simplest lattice-spin systems with vanishing single-ion anisotropy \cite{[1]}. It has been solved theoretically, using various techniques, to observe the meetings of phase boundaries between the ordered and disordered states \cite{[2],[3],[4],[5],[6],[7],[8],[9],[10],[11],[12],[13],[14],[15],[16],[17]}. In the prevailing majority of studies, calculations were carried out using mean-field theory (MFT). In the zeroth- and first-order approximations, the MFT self-consistent expressions for the dipolar and quadrupolar order parameters were derived using the free energy minimization for the spin system at equilibrium. Solutions of these expressions give various branches including stable, metastable and unstable states of the system \cite{[2], [9],[10],[11],[12],[14]}. The MFT calculations of the isotropic BEG model from Tanaka and Mannari work \cite{[2]} were combined with phenomenological and statistical nonequilibrium theories by Erdem and co-workers in order to determine the thermal dependence of several dynamic quantities \cite{[18],[19],[20],[21],[22],[23]}.

A similar approach based on the MFT self-consistency has recently been employed to calculate the curvature scalar or Ricci scalar ($R$) of Ising magnets with two spin states by one of us \cite{[24],[25],[26]}. In these references, for a given manifold $\mathcal{M}$ defined in the thermodynamic state space, a metric called Ruppeiner metric \cite{[27],[28]} was found from the Hessian matrix (HM) of mean-field free energy. Because the derivatives in the HM are evaluated in the equilibrium state, Ricci scalar was derived in terms of the known equilibrium values of order parametres. Among the conclusions, it was noted that in the whole range of external parameters, the curvature scalar remains positive ($R > 0$) and thus no sign change in  $R$ occurs. Another evalution of  $R$ was Ref. \cite{[29]} where it was shown by Mirza and Talaei that the two-dimensional Ising model on a Kagome lattice in a magnetic field has diverging curvature to $\pm \infty$ on opposite sides of the phase transition line ($R<0$  on the ferromagnetic ($F$) side, and $R > 0$ on the frustrated side). Likewise, $R$  has already been worked out extensively in a number of discrete spin systems \cite{[30],[31],[32],[33]}. Although much work are devoted to thermodynamic geometry of magnetic systems, the sign of $R$ is still less explored in most spin systems than in fluid systems.

In the present work, we extend the Ruppeiner's geometrical treatment of thermodynamics to the isotropic BEG model. This extension is more interesting than abovementioned geometric formulations of Ising magnets, because $R$ for the present spin system is negative on the $F$ side of continuous phase transition line with a divergence to $- \infty$  near the zero temperature but it diverges to $+ \infty$  below the critical temperature. Conversely, $R > 0$ on the paramagnetic ($P$) side. The vanishing curvature line ($R=0$) which separates the $R > 0$  and $R < 0$ phase regimes will be determined in the phase diagram and the results will be analyzed in comparison with the other curvature calculations \cite{[34]}.

This paper is organized as follows: Section 2 contains a brief recall presentation of isotropic BEG model and its equilibrium properties under the MFT. In Section 3, we deal with the derivation of Ricci scalar using Ruppeiner metric. The behavior of  $R$  near the phase transition temperatures is shown in Section 4. Finally, some concluding remarks are given in the last section.

\section{The model and mean-field solutions for the system at equilibrium}

For a system of $N$ spins, the Hamiltonian of the isotropic BEG model is given by \cite{[2],[3],[4],[5],[6],[7],[8],[9],[10],[11],[12],[13],[14],[15],[16],[17],[18],[19],[20],[21]}
\begin{equation}
\mathcal{H}= - J \sum_{<ij>} S_i S_j - K \sum_{<ij>} S_i^2 S_j^2, \label{ham}
\end{equation}
where $S_i$ is the spin variable (at site $i$) which take on three values $0, \pm 1$ and the interactions are between  nearest-neighbor pairs $<ij>$ only. $J$ and $K$ are the bilinear and biquadratic exchange coupling energies, respectively. The properties of the system at equilibrium are generally determined self-consistently using Gibbs energy calculations. Letting $U$, $T$, $\sigma$, $H$, $D$, $m$, and $q$ be the internal energy, absolute temperature, entropy per site, external magnetic field, single-ion anisotropy, dipolar order parameter (or magnetization per site), and quadrupolar order parameter (or quadrupole moment per site), respectively, the Gibbs energy functional is given by the following general formula \citep{[35]}
\begin{equation}
G = U - T N \sigma - H N m + D N q.   
\end{equation}
Here, we assume $m \equiv <S_i>$ and $q \equiv <S_j^2> - 2/3$ where $<...>$ denotes the thermal expectation value. We recall that Eq. (2) represents the total Legendre transform of entropy with respect to the variables $m$ and $q$. Under the MFT in the Bragg-Williams formalism, it can be rewritten in the form    
\begin{equation}
\phi( m, q) =  \frac{1}{K}\frac{G}{N} = - r m^2 - q^2 + \theta \sum_{i=1}^{3} p_{i}\ln p_{i}- h m + d q, \label{f-mq}
\end{equation}
where $r = J/K$ is the coupling ratio constant, $\theta = k_B T / K$  is the reduced temperature ($k_B$ the Boltzmann constant), $h = H/K$, $d=D/K$,   and $p_{i}$ are the probabilities of the spin states consistent with given values of dipolar and quadrupolar order parameters. They are defined by
\begin{equation}
p_{1}=\frac{1}{3}+\frac{1}{2}m+\frac{1}{2}q,\,\,\, p_{2}=\frac{1}{3}-q,  \,\,\, p_{3}=\frac{1}{3}-\frac{1}{2}m+\frac{1}{2}q.  \label{omega-mq}
\end{equation}
The equilibrium values of the dipolar and quadrupolar order parameters are found from the conditions $\partial f / \partial m = 0$  and $\partial f / \partial q = 0$ which yield the following set of self-consistent equations for the isotropic case ($d=0$) 
\begin{equation}
m = \frac{ 2 e^z \sinh(y)}{1 + 2 e^z \cosh (y)}, \,\,  q = \frac{2}{3} \left(  \frac{ e^z \cosh (y) - 1}{1 + 2 e^z \cosh (y)} \right) .   \label{m-q}
\end{equation}
Here, the notations $y = ( 2 r m + h) / \theta $ and $z = 2 q / \theta $ are introduced for simplicity. From the numerical solutions of (\ref{m-q}), the relevant equilibrium states have been investigated for several values of $r$. According to the graphical results shown for $h=0$ by Tanaka and Mannari \cite{[2]}, order parameters $m$  and $q$  start from their saturated values ($m \approx 1$ and $q \approx 1/3$), decrease continuously with increasing $\theta$ and approach to zero at critical temperature ($\theta_{C} = 4 r / 3$) while $r \geqslant 2/3$. This phenomenon is known as a continuous (or second-order) phase transition from the $F$ phase ($m > q > 0$) to $P$ phase ($m = q = 0$). In this case, the order parameters can be expressed in the vicinity of $\theta_{C}$  by
\begin{equation}
m \cong 2 \left( \frac{1}{2r} \frac{3r-1}{3r-2} \right)^{1/2} \delta^{1/2}, \,\,  q \cong \frac{1}{2} \left(  \frac{ 3}{3 r -2} \right) \delta ,  \label{m-q-2}
\end{equation}
where  $\delta= \theta_{C} - \theta$, a measure of distance from the critical temperature. On the other hand, both order parameters jump to zero at the critical point when $1/3 \leqslant r < 2/3$, called a discontinuous (or first-order) phase transition and the system has a tricritical point (TCP) for $r=2/3$. Also, one can observe that a first-order phase transition between quadrupolar ($Q$) ($m=0, q<0$) and $P$ phases exists, if $r < 1/3$. In the presence of an external magnetic field ($h \neq 0$), the phase transition has been removed \cite{[11]}. These solutions are usually known as the stable branch. There exists another set of solutions (or metastable and unstable branches) which can obviously be seen in Refs. \cite{[2],[9],[10],[11],[12],[14]}. Above information is very important for the geometrical analysis of critical  and tricritical properties, particularly for the investigation of Ricci scalar vs. reduced temperature plots.

\section{Ruppeiner geometry and derivation of Ricci scalar for the isotropic BEG model}

Now, considering a $n$-dimensional thermodynamic state space (manifold $\mathcal{M}$), a metric (line element) is defined as follows:
\begin{equation}
ds^2 = g_{ij} dx^i dx^j, \qquad ( i,j = 1,2,...,n), \label{metric}
\end{equation}
where $x^i$ denotes the various thermodynamic variables. In the Ruppeiner approach \citep{[27]}, components of the metric tensor are usually found from the Hessian of the entropy with respect to the extensive quantities of a system. As indicated in the third line of Table II in Ref. [28], the total Legendre transform of  entropy or internal energy with respect to variables $x^i$ is another convenient quantity to calculate the metric elements via the definition     
\begin{equation}
g_{i j} = -\beta \partial_i \partial_j \phi\, ,\label{metriccomponet}
\end{equation}
where $\phi$ is the thermodynamic potential per site, $\beta = 1/k_B T$ and $\partial_i = \partial / \partial x^i $. In terms of the metric elements $ g_{ij}$, the Christoffel symbols ($\Gamma^{i}_{jk}$) and curvature tensor ($R^{i}_{jkl}$) are, respectively written. Then, the Ricci tensor is defined by $R_{ij}=R^{k}_{ikj}$, and after another contraction of the Ricci tensor indexes, follows the Ricci scalar
\begin{eqnarray}
& & R = g^{ij}R_{ij}, \qquad  \label{ricci}
\end{eqnarray}
where $g^{ij}$ are the components of contravariant metric tensor. Eq. (\ref{ricci}) is also called as the thermodynamic curvature which  measures the complexity of a system and plays a central role in any attempt to look at phase transitions from geometrical perspective. Specifically, the sign of $R$ indicates whether the interactions in fluids and spin systems are effectively attractive ($R<0$) or repulsive ($R<0$). In the literature, it has been calculated and analysed for a variety of critical phenomena \cite{[34],[36],[37],[38]}.

For the spin system under consideration, we firstly parametrize a two-dimensional manifold by $( x^1,x^2) = (m,q)$. Then, one can find the elements of a nondiagonal metric tensor directly using (\ref{f-mq}) and (8) as follows:
\begin{equation}
g_{11} = - \frac{1}{\theta} \frac{ \partial^2 \phi}{\partial m^2}, \,\, g_{12} = - \frac{1}{\theta} \frac{ \partial^2 \phi}{\partial m \partial q}, \,\, g_{22} = - \frac{1}{\theta} \frac{ \partial^2 \phi}{\partial q^2} , \label{metric-coef}
\end{equation}
where the derivatives are evaluated in the equilibrium state. Using the results of (\ref{metric-coef}) in (\ref{ricci}) Ricci scalar is found in terms of the known equilibrium values of $m$ and $q$ determined by the self-consistent equations (\ref{m-q}) as follows:
\begin{equation}
R (m,q) = - \frac{81}{2}  \frac{A}{B^{2}}  \theta^3 \, , \label{ricci-2}
\end{equation}
where the coefficients $A$ and $B$ are defined by
\begin{equation}
A = 18 (r m^2 + q^2) + 12 q (3 r-1) - 9 \theta + 6 r + 2 \, , \label{ricci-A}
\end{equation}
\begin{eqnarray}
& & B = - 108 r q (m^2 - q^2 -q) - 54 \theta ( r m^2 + q^2 - r q)  \nonumber \\   & & \,\,+ 36 r( m^2 + \theta) - 27 \theta^2 - 18 q \theta + 12 \theta - 16 r \,. \label{ricci-B}
\end{eqnarray}
Some examples of numerical calculations related with Eq. (\ref{ricci-2}) and the discussion of the results will be given in the next section.

\section{Results and discussion}

The variation of Ricci scalar $R$ as a function of the reduced temperature ($\theta$)  in the $F$, $P$ and $Q$ phases is shown in Figs. 1(a)-1(c) for several values of $r$, which correspond to the second-order phase transition temperature, the first-order phase transition temperature, and the TCP. In the figures, the vertical dotted lines refer to phase transition temperatures when there is no external magnetic field and the horizontal dotted lines are for $R=0$ case. According to results displayed in Fig. 1(a), Ricci scalar for $r \geqslant 2/3$ is mostly negative and dependent on $r$ at low temperatures in the $F$ phase. Also, it decreases and $R\rightarrow - \infty$ as the temperature is lowered to zero which is a novel result and not found in earlier studies. This property can be verified by inserting the saturated values of the order parameters ($m=1.0$  and $q=1/3$) into Eqs. (12) and (13) and resulting expressions to (11). Thus, we observe that $R\propto -\theta^{-1}$ corresponding to a tendency to $-\infty$ as $\theta \rightarrow 0$.  On the other hand, the Ricci scalar $R$ decreases with increasing temperature but it reaches to a minimum value ($R_M \approx - 69.255$) when $\theta_M \approx 0.877$ for $r=2/3$. Further increasing of  $\theta$  causes a sign change at $\theta_{SC} \approx 0.882$. Then a rapid increase or a tendency to the plus infinity ($R\rightarrow +\infty$) near the temperature of tricritical point ($\theta_{TCP}\approx 0.888$) on both sides in the $R>0$ region is observed (see the red curve in Fig. 1(a)). The same picture is also displayed when $r=1.2$, illustrated by the blue curve in Fig. 1(a). In this case, the minimum in the $F$ phase (with $\theta_M \approx 1.493$, $R_M \approx -11.50$) is wider and less deeper than that in the tricritical one. After changing its sign from negative to positive at $\theta_{SC} \approx 1.541$ a very rapid divergence to positive infinity occurs around $\theta_C = 1.6$. In order to analytically investigate the behavior of the Ricci scalar just below $\theta_C $, we insert (6) into (12) and (13) to obtain
\begin{equation}
R (\delta) = \frac{X}{4Y} \delta^{-2} \, , \label{ricci-3}
\end{equation}
where $X=\sum_{k=0}^4 a_k (r) \delta^k$ and $Y=\sum_{\ell = 0}^4 b_{\ell} (r) \delta^{\ell}$. From Eq.   (\ref{ricci-3}), one can conclude that when $r$  is a finite number with $r > 2/3$, corresponding to the cases $a_0 (r) > 0$ and $b_0 (r) > 0$, $R$ is always positive and tends to plus infinity as $\delta \rightarrow 0$ or $\theta \rightarrow \theta_{C}$ from below. Using again Eq. (\ref{ricci-3}), it is found that the Ricci scalar just below the criticality is expressed as $R(\delta) \propto \delta^{\lambda}$ with a curvature exponent of $\lambda= -2$. It has long been argued that near a critical point $R(\delta) \propto \delta^{\alpha -2}$ where $\alpha$ is the specific heat exponent with $\alpha \geqslant 0$ \cite{[29]}. In addition, most of the phase transitions observed in the framework of BEG model is represented by the discontinuous jumps of specific heat which correspond to $\alpha =0$ \cite{[39]}. Hence, setting $\alpha =0$ we reach the same result $R(\delta) \propto \delta^{-2}$. The value of curvature exponent reported in this study ($\lambda = -2$) matches exactly those of other spin models with $d \geqslant 4$ dimensions although there is no physical relation between them. Therefore, above results provide a similar example of statistical approximation in which the curvature of thermodynamic metric diverges at the critical point.

Another critical behavior is the large jump of Ricci scalar ($\Delta R \approx 10000$) at the first-order phase transition temperature where $R$ also changes its sign abruptly from  $R < 0$ in the $F$ phase to $R > 0$ in the $P$ phase, seen in Fig. 1(b). Similarly, going from  $Q$ phase to $P$ phase by raising the temperature at a constant $r$ value gives a first-order phase transition accompanied by an abrupt increase with a sign change in $R$ (Fig. 1(c)). The only difference in both figures is that Ricci scalar is independent of $r$  in the whole range of temperature in the $Q$ phase while it slightly depends on $r$  just above the phase transition temperature and becomes again  $r$-independent at very high $\theta$  in the $P$ phase. Also in this category is the ferroelectric crystals. In this family of crystals, $R$ is negative for lower energy gap ($\epsilon$) and pressure ($p$) values and sharply increases to large positive values at the first-order ferroelectric-paraelectric phase transition as either $\epsilon$ or $p$ is increased [34].

\begin{figure*}
\centering
\includegraphics[width=0.45\linewidth]{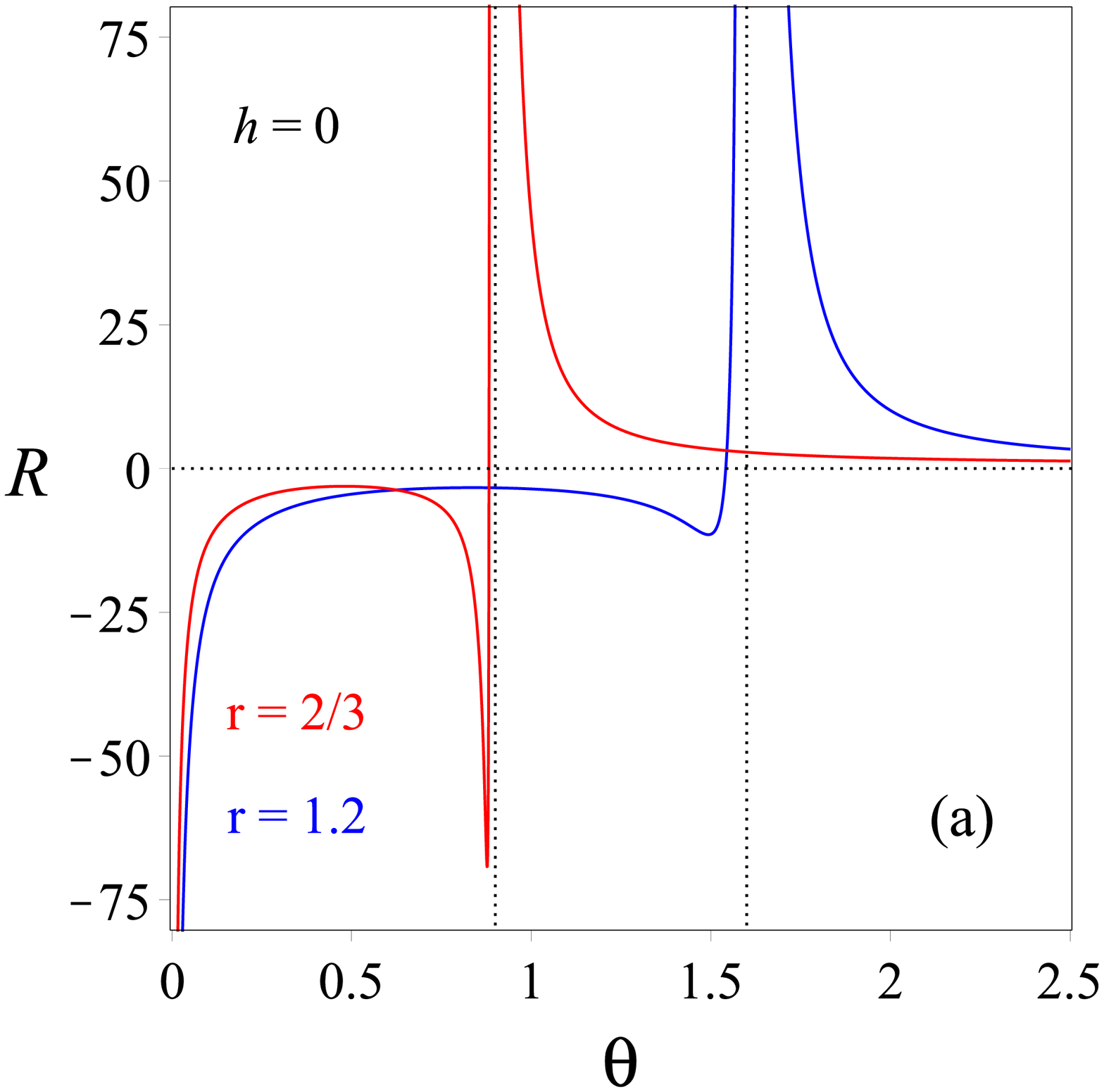}
\includegraphics[width=0.45\linewidth]{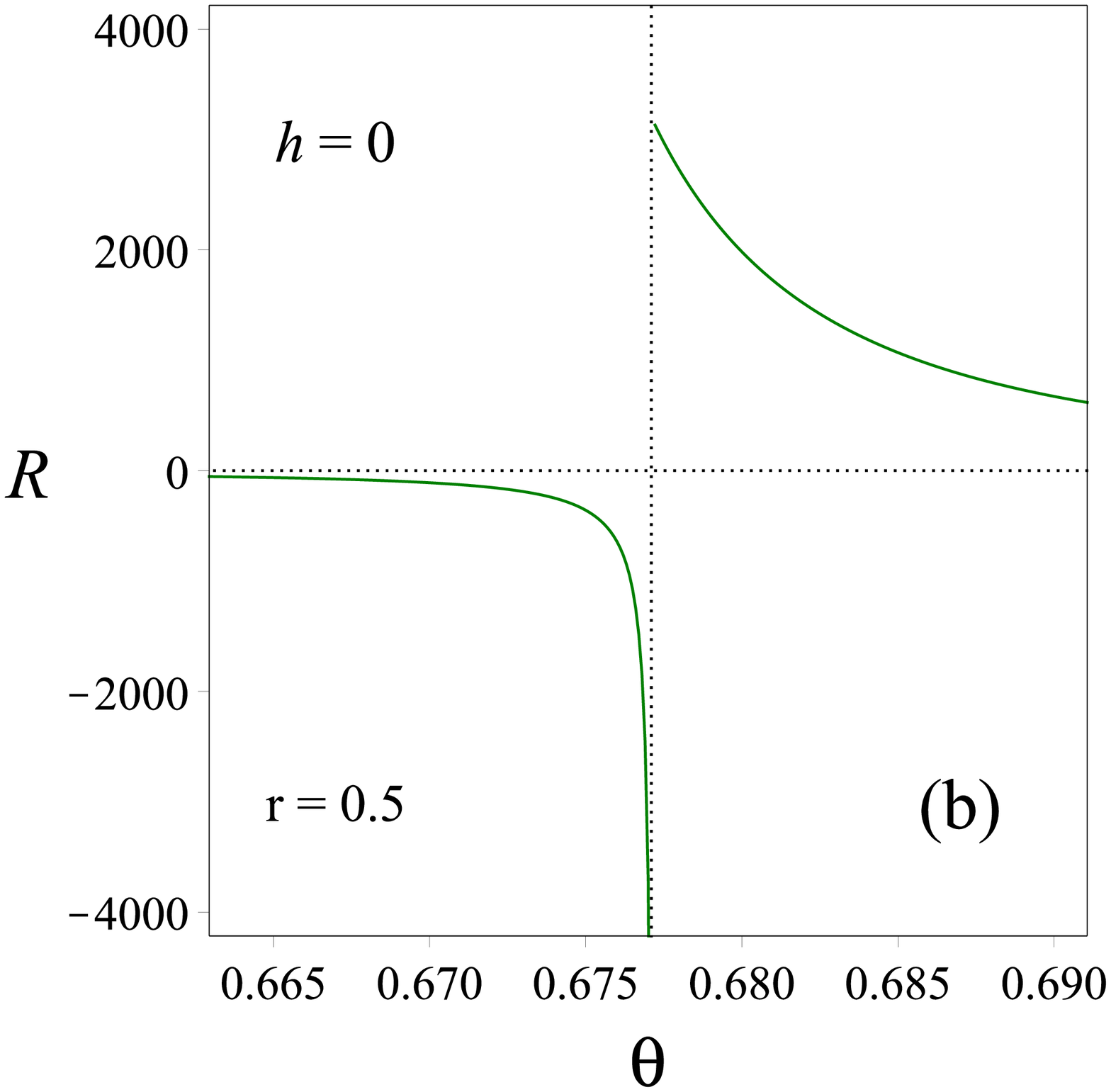}
\includegraphics[width=0.45\linewidth]{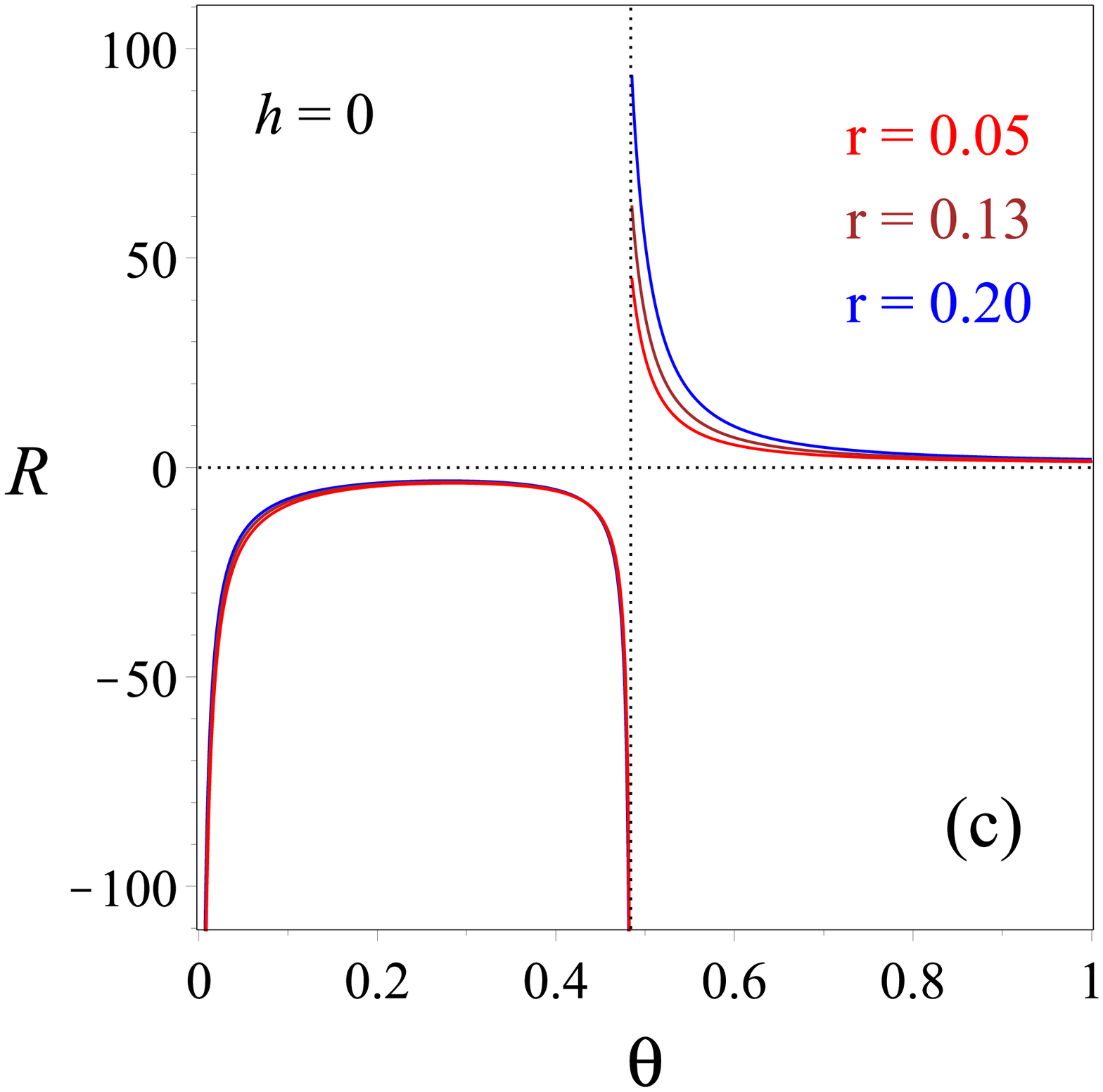}
\includegraphics[width=0.45\linewidth]{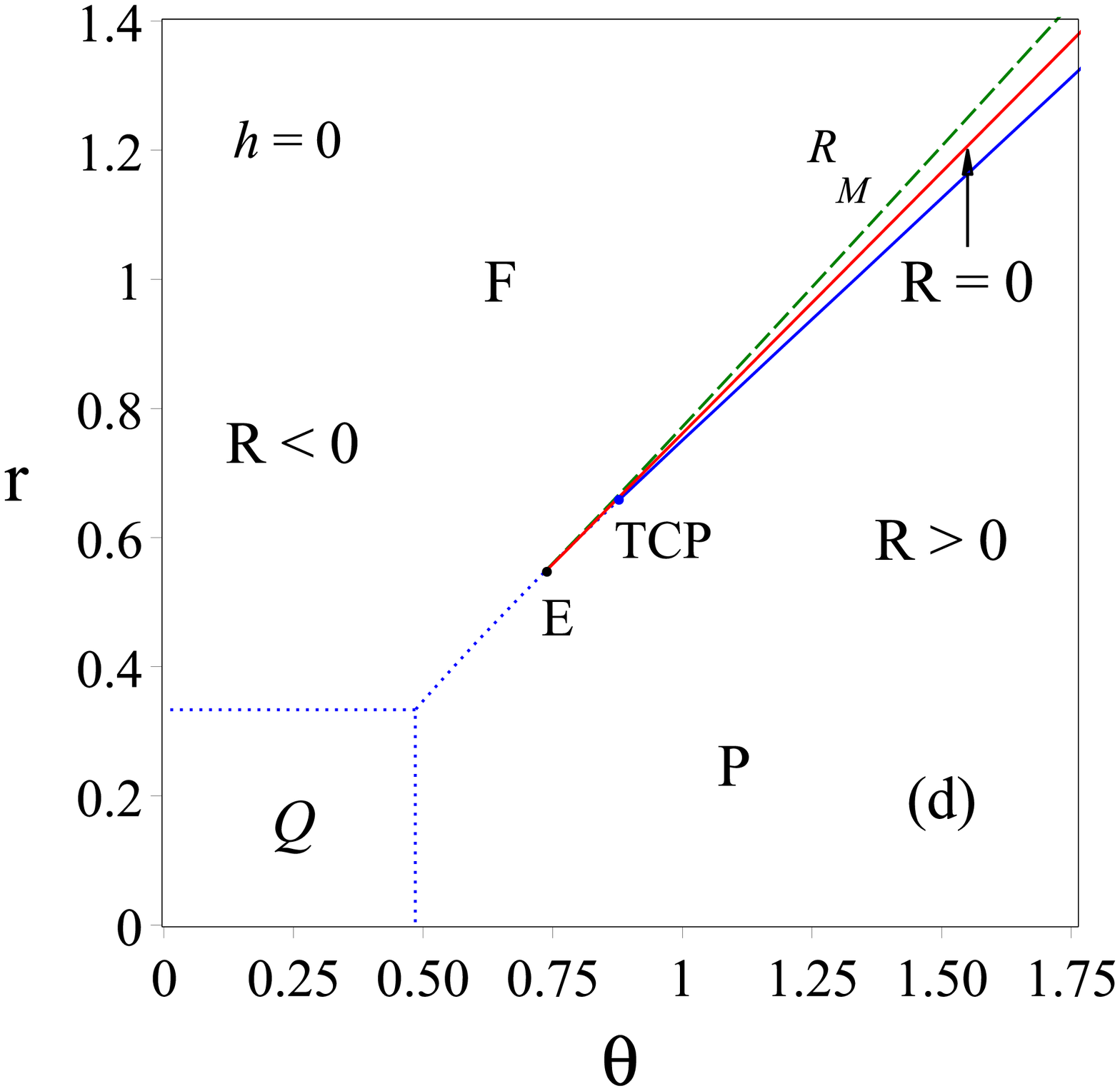}
\caption{ (a) Ricci scalar $R$ vs. reduced temperature $\theta$ for $r\geqslant 2 /3$. (b) Same as Fig. 1(a) but for $1/3 \leqslant r < 2/3$. (c) Same as Fig. 1(a) but for $r < 1/3$. (d) Geometric phase diagram with a $R=0$ boundary line in the $r-\theta$ plane. $h=0$.}
\label{fig1}
\end{figure*}

Besides the study of $R$ along the $F/P$ phase equilibria, we now locate the $R=0$ boundary line in the phase space, which is very close to conventional phase transition indicated by the blue lines in Fig. 1(d). Across the red line in the figure, $R$ changes its sign smoothly from negative to positive as one increases the temperature. This change is mostly referred as geometric phase transition [40]. Unlike the phase boundary associated with singularity in $R$, the geometric phase transitions from $R<0$ phase to $R>0$ phase are not associated with any divergence behavior. In the geometric phase diagram, both $R=0$ line and the line of $R_{M}$ terminate at the same point ($E$) ($r_E\approx0.55$, $\theta_E\approx 0.739$) located on the discontinuous phase transition line (dotted blue line). This is very similar to the critical end-point definition presented by Hoston and Berker [41]. 

\begin{figure*}
\centering
\includegraphics[width=0.45\linewidth]{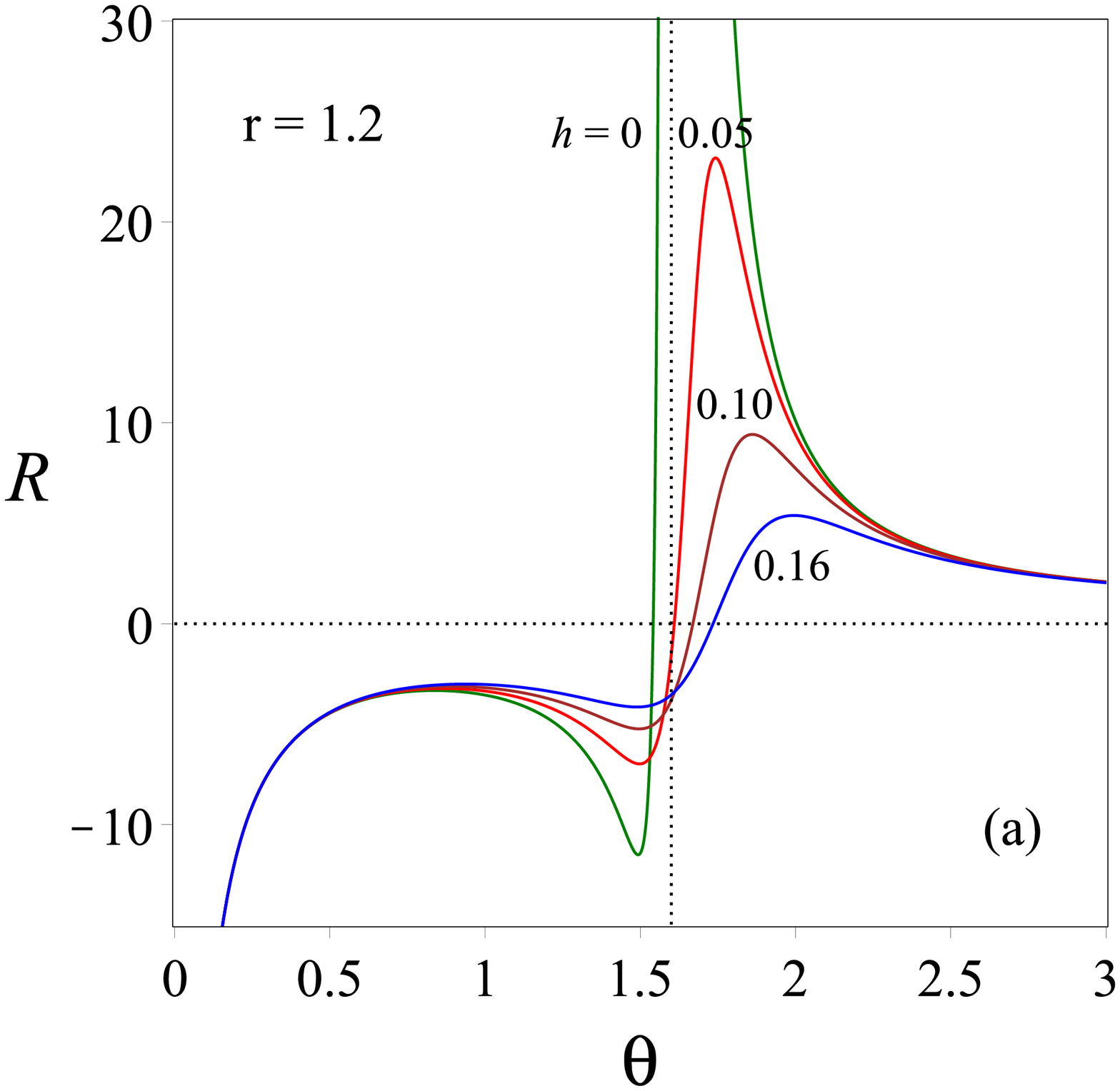}
\includegraphics[width=0.45\linewidth]{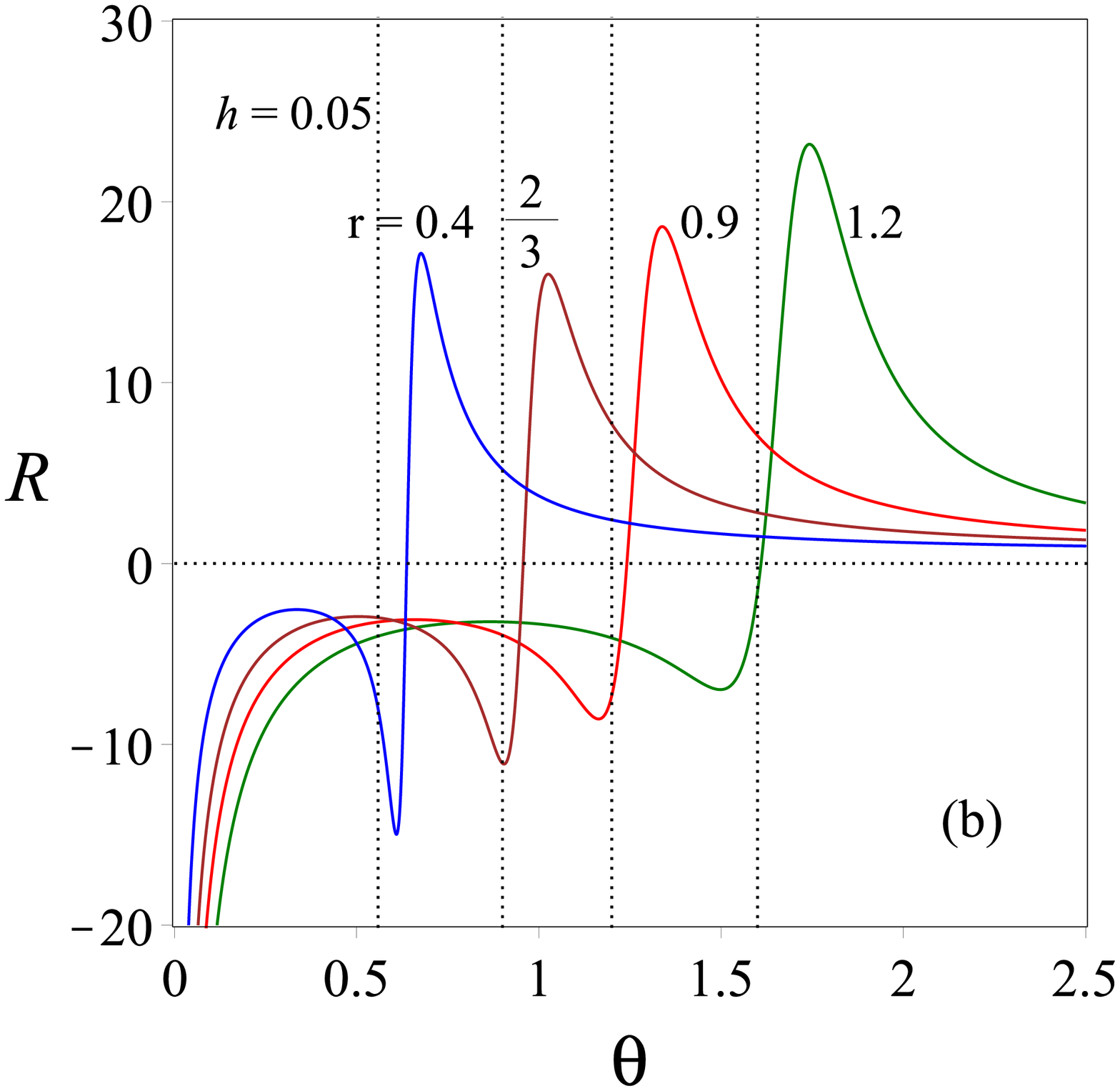}
\includegraphics[width=0.45\linewidth]{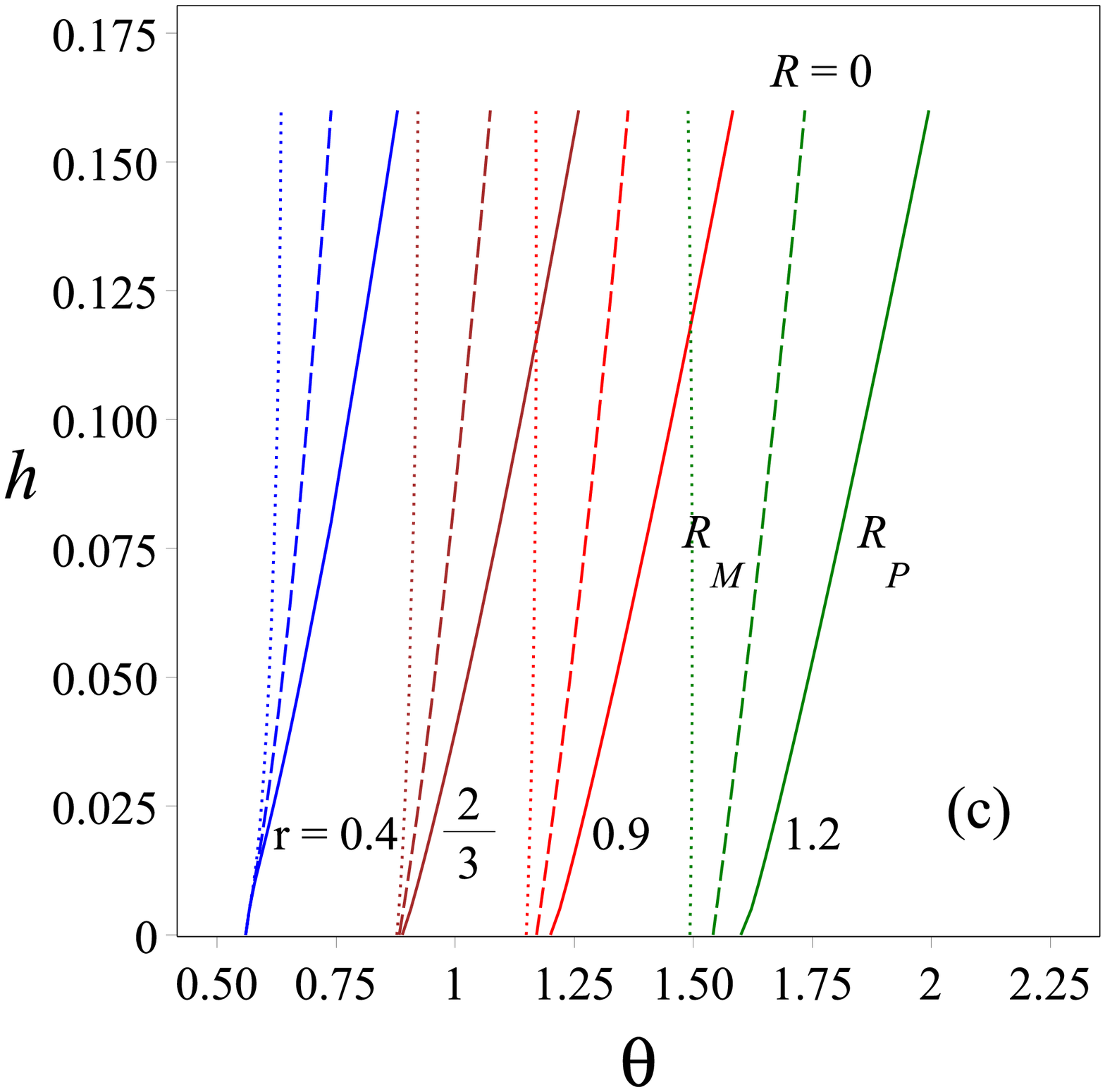}
\includegraphics[width=0.45\linewidth]{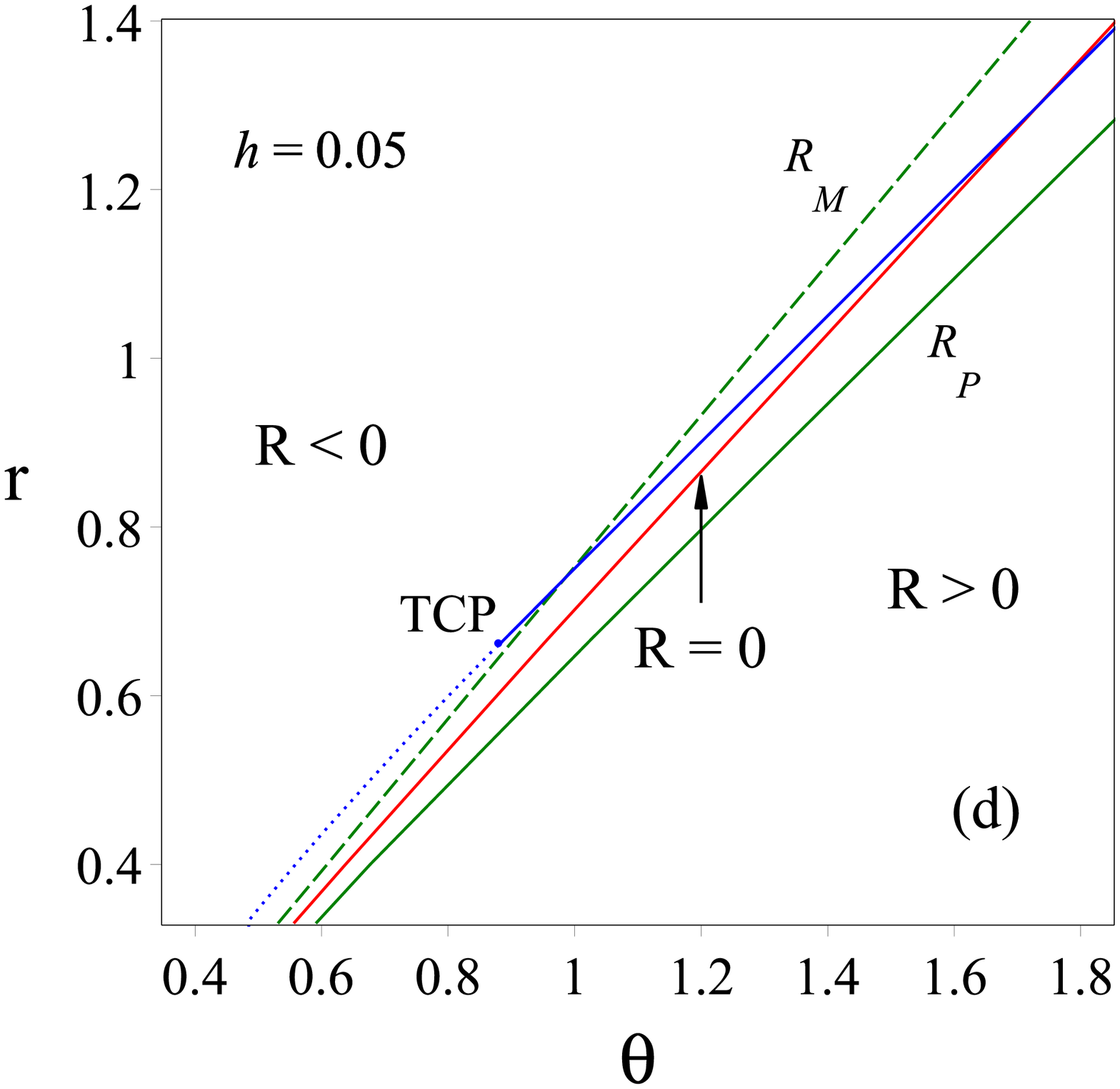}
\caption{(a) Ricci scalar $R$ vs. reduced temperature $\theta$ for the case $h \neq 0$ at $r=1.2$. (b) Same as Fig. 2(a) but for the case $\alpha \geqslant 1/3$ at $h=0.05$. (c) Loci of $R_P$ and $R_M$ with $R=0$ line for several values of $r$ in the $h-\theta$ plane. (d) Same as Fig. 2(c) but in the $r-\theta$ plane when $h=0.05$.} \label{fig2}
\end{figure*}

In the next two figures (Figs. 2(a) and (b)), we show different types of thermal variation of Ricci scalar for $h \neq 0$ with the other parameters taken as in Fig. 1(a). As can be seen, there is no anomaly of $R$ because external magnetic field cancels the singularities observed in Fig. 1. The existence of smooth curves with two new characteristics is illustrated: one with large maximum or a peak ($R_{P}$) above the critical point, and another with a small minimum ($R_{M}$) below $\theta_{C}$. The shapes and positions of these extrema depend both on the magnetic field and exchange coupling ratio. When  $r$ is fixed ($r=1.2$) but $h$ increases, the extrema become smaller and shift towards higher temperatures (Fig. 2(a)). But, on increasing $r$ at fixed $h$ ($h=0.05$) $R_{P}$ ($R_{M}$) becomes larger (smaller) while the same shifts occur (Fig. 2(b)).

On the other hand, the extrema $R_{P}$ of the curves observed in Fig. 2(a) determine the line of curvature maxima thus indicating the loci of correlation length maxima or the Widom line for the isotropic BEG model. In Fig. 2(c), we construct four Widom lines (colored solid curves) predicted for $r=0.4$, $2/3$, $0.9$, $1.2$. They start at their critical temperatures indicated by the vertical dotted lines in Fig. 2(b) and extend into the region $h>0$ in $h-\theta$ plane illustrated in Fig. 2(c). We compare our results for Widom lines with those of the 1D Ising model on the $h-\theta$ plane where a manifold $\mathcal{M}$ with $(x^1,x^2)=(\beta J, h$) is used by Dey et al. [32]. We find a good agreement between the spin models, thus supporting the power of ($m,q$) choice in the geometrical investigations. It is natural to ask whether the $R_{P}$ line might coincide with the $R = 0$ line (colored dashed curves) and $R_{M}$ line (colored dotted curves) since these lines mark a transition from atractive to repulsive interactions in fluid systems \cite{[36]}. We proceed to analyze our results comparing above curves for various values of $r$ in Fig. 2(c). Indeed, three curves join at the temperature of $\theta_{TCP}$ when $h=0$, $r=2/3$ and coincide only for very small $h$ values if $r<2/3$. This fact represents  a possible correspondence between $R=0$ and Widom curves (For an example of such a correspondence in fluids see [36]). Finally, loci of extrema of $R$ can also be worked out in the full range of $r$ with $r \geqslant 1/3$. For a special value of $h$ ($h=0.05$), determinations of $R_{P}$ and $R_{M}$ are straightforward and the results are represented in Fig. 2(d). It is shown that both $R_{P}$ (green solid line) and $R_{M}$ (green dashed line) extend into the region $r \geqslant 1/3$ in $h-\theta$ plane on opposite sides of $R=0$ boundary (red solid line). Note that the conventional phase boundary lines from Fig. 1(d) (blue solid and dotted lines) have also been replotted into Fig. 2(d) so that the reader can compare them with the results of $R=0$ boundary line.

\section{Conclusions}

As a conclusion, we have reported here on the microscopic basis for a geometrical extension of the isotropic BEG model. This is the first and novel treatment of geometrical approach in spin-1 Ising systems. From the results for the Ricci scalar presented in Fig. 1, the spin system has two different behaviors depending on whether $\theta < \theta_{C}$  or  $\theta > \theta_{C}$. For $\theta < \theta_{C}$, the Ricci scalar $R$ is mainly negative with a divergence singularity at $\theta = 0$ ($R\rightarrow - \infty$). Thus, we may call the ferromagnetic spin interaction is attractive, while for $\theta > \theta_{C}$, $R$ is always positive and so the interaction between the spins is repulsive, as in other solids [42]. We have shown that there are $r$-dependent critical values for the reduced temperature $\theta_{SC}(r)$ where $R$ varies rapidly and undergoes a change of sign. Such a geometric transitions are necessarily continuous. We also confirmed numerically theoretical predictions for the critical point properties of $R$ and find that the curvature scalar diverges with an exponent of $-2$ below $\theta_{C}$. We notice that these findings are consistent with the recently developed interpretation that Ricci scalar gives for the ordered/disordered phases in the ferroelectric crystals. This consistency is an expected result, since, as identified in Ref. [34], the quantum lattice version of the BEG model corresponds precisely to Eq. (1) with $H=0$. After introducing the external magnetic field, $R$ becomes smooth functions of $\theta$ with broad extrema in the $R<0$ and $R>0$ regimes (Fig. 2). In this case, $\theta_{SC}$ depends on both $r$ and $h$. Another issue is whether above features exist beyond two-dimensional ($n=2$) manifolds that we have studied here. In this regard, we conclude by noting that preliminary results for a similar analysis of three-dimensional manifold where $(x^1,x^2,x^3)=(m,q,\theta$) also clearly indicate the presence of a negative curvature region, although this domain in parameter space is somewhat smaller than that for $n=2$ case mentioned above.

\begin{acknowledgements}
We would like to thank Prof. G. Ruppeiner (New College of Florida, USA) for useful discussions related with the topic.
\end{acknowledgements}

\end{document}